\documentclass[pra,superscriptaddress,showpacs,preprintnumbers,eqsecnum,amsmath,amssymb]{revtex4}

\begin{document}

\title{Geometrically Underpinned Maximally Entangled States Bases}

\author{M. Revzen}
\affiliation {Department of Physics, Technion - Israel Institute of Technology, Haifa
32000, Israel}

\date{\today}

\begin{abstract}
Finite geometry is used to underpin finite, two d-dimensional particles Hilbert space,
d=prime $\ne2$. A central role is allotted to states with mutual unbiased bases (MUB)
labeling. Dual affine plane geometry (DAPG) {\it points} underpin single particle, MUB labeled,
product states. The DAPG {\it  lines} are shown to underpin maximally entangled states which form
an orthonormal basis spanning the space. The relevance of mutually unbiased collective
coordinates bases (MUCB) for dealing with maximally entangled states is discussed and shown
to provide an economic alternative mode of study. These maximally entangled, geometrically
reasoned states, provide the resource to a transparent solution to what may be termed tracking  of the Mean King Problem (MKP): here Alice prepares a state measured by King  along some orientation which Alice succeed in identifying  with a subsequent measurement. Brief expositions of the topics considered: MUB, DAPG, MUCB and the MKP are
included, rendering the paper self contained.
\end{abstract}


\maketitle

\section {  Introduction}

Several recent studies
\cite{bengtsson,bengtsson1,berge2,vourdas3,saniga,planat1,planat2,combescure,wootters4}
consider the affinity of finite, d, dimensional Hilbert space to finite Galois fields,
GF(d), and thereby to finite geometry. These interrelations are of interest as they
illuminate both subjects. The present work contains a novel intuitive geometrical
underpinning for the MUB structure of $d^2$ dimensional Hilbert space accommodating two
d-dimensional particles. (d=prime ($\ne 2$).) The study gives for the first time, to our
knowledge, explicit formulae that relates lines and points of the geometry to {\it states}
(rather than projectors) \cite{rev1}
allowing a geometric view of the relation between product and maximally entangled states \cite{ent}.\\
The analysis underpins Hilbert space states (and operators) with geometrical points and
lines. In particular we introduce, in section IV, a
 simple, i.e. universal, balancing term. This term,  denoted by
${\cal{R}}$, arises upon the association of states addition in Hilbert space with
geometrical requirements among points and lines.( The corresponding term  is the unit operator in  studies, \cite{wootters4,rev2}, wherein the association was to Hilbert space projectors \cite{wootters4,rev1}.) Its origin, given in the uncanny
affinity  of mutual unbiased bases (MUB)
 labeling for a  geometrical coordination scheme, is outlined in section IV. Included in this section is a
 brief exposition of  of DAPG  which we use to underpin   a d-dimensional single particle Hilbert space  state projectors
 \cite{wootters4,berge2,vourdas,klimov2,rev1} the results of which are used in the present analysis that pertains to a $d^2$ dimensional Hilbert spaces.

 In section V we give the
 central result of this paper i.e. the demonstration that the state underpinned with
 geometrical line is a maximally entangled state of remarkable attribute: its overlap with (judicially defined to relate to one coordinate point) two particles product  state is definitive. Thus 
 it is 1/d
 if the underpinning point
 is on the line, nil otherwise.  Since  d lines share a point  this relates to d lines.  This holds while the constituent single particle state projectors
 have nonvanishing overlap with each of the $d^2$  (orthogonal maximally) entangled states that span the space.  This  issue  is elaborated on  in section VIII  and  leads to a novel tracking of  
 the Mean King problem outlined in section IX : Alice produces a state which allow the tracking of the alignment of the King apparatus used in his measurement of the state by one subsequent measurement.\\
An alternative approach to the construction of a $d^2$ dimensional maximally entangled states basis is given in section VII where these states are shown to be 
{\it product} states of {\it collective}  coordinates MUB bases. The theory of this aproach is outlined in sections II and III. The account via the collective MUB states proves to be more economic  and, perhaps, more informative physically.

\section{   Finite dimensional Mutual Unbiased Bases, (MUB): Brief Review}

In a d-dimensional Hilbert space, two complete, orthonormal vectorial bases, ${\cal
B}_1,\;{\cal B}_2$,
 are said to be MUB if and only if (${\cal B}_1\ne {\cal B}_2)$

\begin{equation}
\forall |u\rangle,\;|v \rangle\; \epsilon \;{\cal B}_1,\;{\cal B}_2 \;resp.,\;\;|\langle
u|v\rangle|=1/\sqrt{d}.
\end{equation}
The physical meaning of this is that knowledge that a system is in a particular state in
one basis implies complete ignorance of its state in the other basis.\\
Ivanovich \cite{ivanovich} proved that there are at most d+1 MUB, pairwise, in a
d-dimensional Hilbert space and gave an explicit formulae for the d+1 bases in the case of
d=p (prime number). Wootters and Fields \cite{wootters2} constructed such d+1 bases for
$d=p^m$ with m an integer. Variety of methods for construction of the d+1 bases for $d=p^m$
are now available
\cite{tal,wootters3,klimov2,vourdas}. Our present study is confined to $d=p\;\ne2$.\\
 We now give explicitly the MUB states in conjunction with the algebraically complete
 operators \cite{schwinger,amir} set:
 $\hat{Z},\hat{X}$.  Thus we label the d orthonormal states spanning the Hilbert space,
 termed
 the computational basis, by $|n\rangle,\;\;n=0,1,..d-1; |n+d\rangle=|n\rangle$
\begin{equation}
\hat{Z}|n\rangle=\omega^{n}|n\rangle;\;\hat{X}|n\rangle=|n+1\rangle,\;\omega=e^{i2\pi/d}.
\end{equation}
The d states in each of the d+1 MUB bases \cite{tal,amir} are the states of the
computational basis and the d bases:
\begin{equation} \label{mxel}
|m;b\rangle=\frac{1}{\sqrt
d}\sum_0^{d-1}\omega^{\frac{b}{2}n(n-1)-nm}|n\rangle;\;\;b,m=0,1,..d-1.
\end{equation}
Here the d sets labeled by b are the bases and the m labels the states within a basis. We
have \cite{tal}
\begin{equation}\label{tal1}
\hat{X}\hat{Z}^b|m;b\rangle=\omega^m|m;b\rangle.
\end{equation}
For later reference we shall refer to the computational basis (CB) by $b=\ddot{0}$. Thus
the above gives d+1 bases, $b=\ddot{0},0,1,...d-1$ with the total number of states d(d+1)
grouped in d+1 sets each of d states. We have of course,
\begin{equation}\label{mub}
\langle m;b|m';b\rangle=\delta_{m,m'};\;\;|\langle m;b|m';b'\rangle|=\frac{1}{\sqrt d},
\;\;b\ne b'.
\end{equation}
The MUB set is closed under complex conjugation:
\begin{equation}\label{cc}
\langle n|m,b\rangle^{\ast} =\langle n|
\tilde{m},\tilde{b}\rangle,\;\;\Rightarrow\;|\tilde{m},\tilde{b}\rangle = |d-m,d-b\rangle,
\end{equation}
 as can be verified by inspection of Eq.(\ref{mxel}).   (We denote ($m,\ddot{0})$ by $\ddot{m}$ when no confusion should arise.)

This completes our discusion of single particle MUB.

\section{ MUB for Collective Coordinates}

Several studies \cite{durt1,klimov1,fivel1,berge2,ent} consider the entanglement of two
d-dimensional particles Hilbert space via MUB state labeling. We shall now outline briefly
the approach adopted by \cite{ent} that will be used in later sections.\\
The Hilbert space is spanned by the single particle computational bases,
$|n\rangle_1|n'\rangle_2$ (the subscripts denote the particles). These are eigenfunctions
of $\hat{Z}_i$ i=1,2:
$\hat{Z}_i|n\rangle_i=\omega^{n}|n\rangle_i,\;\omega=e^{i\frac{2\pi}{d}}.$ Similarly
$\hat{X}_i|n\rangle_i=|n+1\rangle,\;i=1,2$. We now define our collective coordinates and
collective operators (we remind the reader that the exponents are modular variables, e.g.
1/2 mod[d=7]=(d+1)/2)=4):
\begin{equation}\label{colz}
\bar{Z}_r\equiv \hat{Z}^{1/2}_{1}\hat{Z}^{-1/2}_{2};\;\;\bar{Z}_c\equiv
\hat{Z}^{1/2}_{1}\hat{Z}^{1/2}_{2}.
\end{equation}
Since $\bar{Z}_{s}^{d}=1,\; s=r,c$ we may consider their respective computational
eigen-bases,

\begin{equation}
\bar{Z}_{s}|n\rangle_s = \omega^{n}|n\rangle_s,\;\;n=0,1,...d-1,\;s=r,c.
\end{equation}
clearly  $|n\rangle_r|n'\rangle_c;\;n,n'=0,1,..d-1,$ is a $d^2$ orthonormal basis spanning
the two d-dimensional particles Hilbert space. Eq.(\ref{colz}) implies,
\begin{equation}
\hat{Z}_1=\bar{Z}_r\bar{Z}_c;\;\;\hat{Z}_2=\bar{Z}_r^{-1}\bar{Z}_c
\end{equation}
In a similar fashion we  define the displacement operators,
\begin{equation}\label{colx}
\bar{X}_r\equiv\hat{X}_1\hat{X}_2^{-1};\;\bar{X}_c\equiv\hat{X}_1\hat{X}_2\rightarrow
\hat{X}_1=\bar{X}^{1/2}_r\bar{X}^{1/2}_c,\;\hat{X}_2=\bar{X}^{-1/2}_r\bar{X}^{1/2}_c.
\end{equation}
These entail
\begin{equation}\label{comcol}
\bar{X}_s\bar{Z}_s=\omega\bar{Z}_s\bar{X}_s,\;s=r,c;\;\bar{X}_s\bar{Z}_{s'}=\bar{Z}_{s'}\bar{X}_s,\;s\ne
s'.
\end{equation}
One readily proves \cite{ent},
\begin{equation}\label{relcomdel}
\langle n_1,n_2|n_r,n_c\rangle=\delta_{n_r,(n_1-n_2)/2}\delta_{n_c,(n_1+n_2)/2}.
\end{equation}
Note that the formula is schematic. Thus $|n_1,n'_2\rangle=|n\rangle_1|n'\rangle_2,\;|n_r,n'_c\rangle=
|n\rangle_r|n'\rangle_c$ i.e. they refer to distinct bases.  We have then,
\begin{equation}\label{relcom}
|n_r,n_c\rangle=|n_1,n_2\rangle,\;\;for\;n_r=(n_1-n_2)/2,\;n_c=(n_1+n_2)/2\;\rightleftarrows
n_1=n_r+n_c,\;n_2=n_c-n_r.
\end{equation}
Thence we may consider collective MUB,
\begin{equation}\label{colmub}
|m_s;b_s\rangle=\frac{1}{\sqrt
d}\sum_n^{d-1}\omega^{\frac{b_s}{2}n(n-1)-m_sn}|n\rangle_s,\;m_s,b_s=0,1,...d-1;\;s=r,c.
\end{equation}
Incorporating the respective CB $b_s=\ddot{0}_s$ it is proved in \cite{ent} that the two
d-dimensional particles state,
\begin{equation}\label{ent1}
|m_r;b_r\rangle|m_c;b_c\rangle \;for\;b_r\ne b_c
\end{equation}
is a maximally entangled state. (For $b_r=b_c$ it is a product state for both {\it
particles} and collective coordinates.)  Indeed that it is a maximally entangled state may be seen by tracing out the first particle coordinates,
\begin{equation}
\sum_{n=0}^{d-1} \langle n|_1|\ddot{m}\rangle_c|2m_0\rangle_r \langle 2m_0|_r\langle \ddot{m}|_c|n\rangle_1=\sum_{n,n_r,n_r{'}}\langle n|\ddot{m}+n_r\rangle|\ddot{m}-n_r\rangle_2\omega^{-2n_rm_0}\langle \ddot{m}-n_r{'}|_2\langle \ddot{m}+n_r{'}|n\rangle\omega^{2n_r{'}m_0}=\frac{1}{d}\hat{I}_2.
\end{equation}
where we used Eqs.(\ref{relcom},\ref{colmub}).\\
This completes our review of mutual collective
unbiased bases (MUCB).

 \section{   Finite Geometry and Hilbert Space Operators}

We now briefly review the essential features of finite geometry required for our study
\cite{bennett,shirakova,tomer,wootters4}.\\
A finite plane geometry is a system possessing a finite number of points and lines. There
are two kinds of finite plane geometry: affine and projective. We shall confine ourselves
to affine plane geometry (APG) which is defined as follows. An APG is a non empty set whose
elements are called points. These are grouped in subsets called lines subject to:\\
1. Given any two distinct points there is exactly one line containing both.\\
2. Given a line L and a point S not in L ($S \not\in L$), there exists exactly one line
$L'$ containing S
such that $L \bigcap L'=\varnothing$. This is the parallel postulate.\\
3. There are 3 points that are not collinear.\\
It can be shown \cite{bennett,shirakova} that for $d=p^m$ (a power of prime) APG can be
constructed (our study here is for d=p). Furthermore The existence of APG implies
\cite{bennett,shirakova}the existence of its dual geometry DAPG wherein the points and
lines are interchanged. Since we shall study extensively  DAPG, we list its properties
\cite{shirakova,bennett}. We shall refer to these
by DAPG($\cdot$):\\
a. The number of lines is $d^2$, $L_j,\;j=1,2....d^2.$ The number of points is d(d+1),
$S_{\alpha},\;{\alpha = 1,2,...d(d+1)}.$\\
b. A pair of points on a line determine a line uniquely. Two (distinct) lines share one and
only
one point.\\
c. Each point is common to d lines. Each line contain d+1 points.\\
d. The d(d+1) points may be grouped in sets of d points, no two of a set share a line. Such
a set is designated by $\alpha' \in \{\alpha \cup M_{\alpha}\},\; \alpha'=1,2,...d$.
($M_{\alpha}$ contain all the points not connected to $\alpha$ - they are not connected
among themselves.) i.e. such a set contains d disjoined (among themselves) points. These
are equivalent classes of the geometry \cite{bennett}. There are d+1 such sets:
$$\bigcup_{\alpha=1}^{d(d+1)}S_{\alpha}=\bigcup_{\alpha=1}^d R_{\alpha};\;\;
R_{\alpha}=\bigcup_{\alpha'\epsilon\alpha\cup M_{\alpha}}S_{\alpha'};\;\; R_{\alpha}\bigcap
R_{\alpha'}=\varnothing,\;\alpha\ne\alpha'.$$ e. Each point of a set of disjoint points is
connected to every other point not in its set.\\
DAPG(c) allows the definition, which we adopt, of $S_{\alpha}$ in terms of addition of $L_j$ which acquires a meaning upon viewing the
points ($S_{\alpha}$) and the lines ($L_j$) as underpinning Hilbert space entities (e.g.
projectors or states, to be specified later):
\begin{equation}\label{A}
S_{\alpha}=\frac{1}{d}\sum_{j\in\alpha}^{d} L_j.
\end{equation}
This implies,
\begin{equation}\label{R}
\sum_{\alpha{'} \in \alpha \cup M_{\alpha}}^{d}S_{\alpha'}=\frac{1}{d}\sum^{d^2}_{j} L_j,
\end{equation}

DAPG(d)  via Eq.(\ref{R}) implies,

\begin{equation}\label{I}
{\cal{R}}=\sum_{\alpha' \in \alpha \cup
M_{\alpha}}^{d}S_{\alpha'}=\frac{1}{d}\sum^{d^2}_{j} L_j=
\frac{1}{d+1}\sum^{d(d+1)}_{\alpha}S_{\alpha} \;\;{\it{independent}}\; of\; \alpha.
\end{equation}
This equation, Eq(\ref{I}), reflects relation among equivalent classes within the geometry
\cite{bennett}. It will be referred to as the balance formula: the quantity ${\cal{R}}$
serves as a balancing term. Thus, Eqs.(\ref{A}),(\ref{I}) imply,
\begin{equation}\label{line}
L_j=\sum_{\alpha \in j}^{d+1}S_{\alpha} - {\cal{R}}.
\end{equation}
(Note that in previous studies, \cite{wootters4,rev1}, where the geometrical point $S_{\alpha}$ underpins 
the projector, $S_{\alpha}\rightarrow \hat{A}_{\alpha=m,b}\equiv|m,b\rangle\langle b,m|$ gives ${\cal{R}}=\Bbb{I}$, i.e. independent of $\alpha$.)
 A particular arrangement of lines and points that satisfies DAPG(x),
x=a,b,c,d,e is referred to as a realization of DAPG. We outline in Appendix A the reasoning and proofs for the geometrically based interrelation among the geometrically underpinned Hilbert space operators.\\
This completes our review of finite geometry.

\section{ Realization of DAPG}

We now consider a particular realization of DAPG of dimensionality $d=p \;\ne 2$ which is
the basis of our present study. We arrange the aggregate, the d(d+1) points, $\alpha$, in a
$d\cdot(d+1)$matrix like rectangular array of d rows and d+1 columns. Each column is made
of a set of d points  $R_{\alpha}=\bigcup_{\alpha{'}\epsilon\alpha\cup
M_{\alpha}}S_{\alpha{'}};$  (DAPG(d)). We label the columns by b=$\ddot{0}$,0,1,2,....,d-1
and the rows by m=0,1,2...d-1.( Note that the first column label of $\ddot{0}$ is for
convenience and does not relate to a numerical value. It designates the computational
basis, CB.) Thus $\alpha=m(b)$ denotes a point by its row, m, and its column, b; when b is
allowed to vary - it gives the point's row position in every column defing thereby the line.. We label the left most
column by b=$\ddot{0}$ and with increasing values of b, that relates to the basis
label, we move to the right. Thus the right most column is b=d-1. The top most point in
each column is labeled by m=0
with  m values increasing as one move to lower rows - the bottom row being m=d-1.\\

e.g. for d=3 the underpinning's schematics is:
\[ \left( \begin{array}{ccccc}
m\backslash b&\ddot{0}&0&1&2 \\
0&A_{(0,\ddot{0})}&A_{(0,0)}&A_{(0,1)}&A_{(0,2)}\\
1&A_{(1,\ddot{0})}&A_{(1,0)}&A_{(1,1)}&A_{(1,2)}\\
2&A_{(2,\ddot{0})}&A_{(2,0)}&A_{(2,1)}&A_{(2,2)}\end{array} \right)\].\\

( In the Hilbert space realization of DAPG, A stands for the Hilbert space entity being
underpinned with coordinated point, (m,b). In \cite{rev1} A represented an MUB projector:
$A_{\alpha=(m,b)}=\hat{A}_{\alpha}=|m,b\rangle\langle b,m|$. In the present paper A will be
seen to signify a two particles state to be specified in a subsequent section.) We now
assert that the d+1 points, $m_j(b), b=0,1,2,...d-1,$ and $m_j(\ddot{0})$, that form the
line j which contain the two (specific) points $m(\ddot{0})$ and m(0) is given by (we
forfeit the subscript j - it is implicit),
\begin{equation}\label{m(b)}
m(b)=\frac{b}{2}(2m(\ddot{0})-1)+m(0),\;mod[d]\;\;b\ne \ddot{0};\;m(\ddot{0})=\ddot{m}.
\end{equation}

The rationale for this particular form is
clarified in the next section. Thus a line j is parameterized fully by
$j=(m(\ddot{0}),m(0))$. We now prove that the set $j=1,2,3...d^2$ lines
covered by Eq.(\ref{m(b)}) with the points as defined above forms a realization of DAPG.\\
\noindent 1. Since each of the  parameters, $m(\ddot{0})$ and m(0), can have d values - the
number of lines $d^2$. The number of points in a line is evidently d+1 - one in each
column:   The linearity of the equation precludes having two points with a common value
of b on the same line. DAPG(a).\\
\noindent 2. Consider two points on a given line, $m(b_1),m(b_2);\;b_1\ne b_2$. We have
from Eq.(\ref{m(b)}), ($b\ne \ddot{0},\;b_1 \ne b_2$)
\begin{equation}\label{twopoints}
m(b_i)=\frac{b_i}{2}(2m(\ddot{0})-1)+m(0),\;\;mod[d]\;i=1,2.
\end{equation}
These two equation determine uniquely ({\it for d=p, prime}) $m(\ddot{0})$ and m(0). DAPG(b).\\
\noindent For fixed point, m(b),   $\ddot{m}$  and m(0) are interrelated, $\ddot{m}\Leftrightarrow m(0),$ thus
the number of free parameters is d (the number of points on a fixed column). Thus each
point is common to d lines. That
the line contain d+1 is obvious. DAPG(c).\\
\noindent 3. As is argued in 1 above no line contain two points in the same column (i.e.
with equal b). Thus the d points, $\alpha,$ in a column form a set
$R_{\alpha}=\bigcup_{\alpha'\epsilon\alpha\cup M_{\alpha}}S_{\alpha'},$ with trivially
$R_{\alpha}\bigcap R_{\alpha'}=\varnothing,\;\alpha\ne\alpha',$ and
$\bigcup_{\alpha=1}^{d(d+1)}S_{\alpha}=\bigcup_{\alpha=1}^d R_{\alpha}.$ DAPG(d).\\
\noindent 4. Consider two arbitrary points {\it not} in the same set, $R_{\alpha}$ defined
above: $m(b_1),\;m(b_2)\;\;(b_1\ne b_2).$ The argument of 2 above states that, {\it for
d=p}, there is a unique solution for the two parameters that specify the line containing
these points. DAPG(e).\\
For example the point m(1) is gotten from, Eq.(\ref{m(b)}),
$$ m(1)= \frac{1}{2}(2-1)+2=1\;\;mod[3]\;\;\Rightarrow\;m(1)=(1,1).$$
We illustrate the above for d=3. The line j labeled by  $j=(\ddot{m},m(0))$ is made up of
the 4 points $j:\;1. \big( m(\ddot{0})=(1,\ddot{0});\;2. m(0)=(2,0);\; 3. m(1)=(1,1)$ and
$4. m(2)=(0,2)\big).$ (We shall denote $m(\ddot{0})$ by $\ddot{m}$ when no confusion should
arise.) The bracketed numbers give the point's coordinates.
\[ \left( \begin{array}{ccccc}
m\backslash b&\ddot{0}&0&1&2 \\
0&\cdot&\cdot&\cdot&(0,2)\\
1&(1,\ddot{0})&\cdot&(1,1)&\cdot\\
2&\cdot&(2,0)&\cdot&\cdot\end{array} \right)\].\\

In \cite{rev2,rev1} we considered the DAPG's points as MUB projectors ( the present paper involves the underpinning of product states by the geometrical points):
$$\alpha=(m,b)\;\Rightarrow\;\hat{A}_{m,b}=|m,b\rangle\langle b,m|.$$ This
scheme allows relating the underpinning DAPG lines to interrelation among the Hilbert
space operators (or states) that form those lines as follows. For $b\ne\ddot{0}$ we have, cf.
Eq.(\ref{mxel}),
\begin{equation}\label{nn'}
\langle n|\hat{A}_{m,b}|n'\rangle=\omega^{(n-n')[b/2(n+n'-1)-m]}/d.
\end{equation}
Thus for fixed b, $(\ne \ddot{0}),$ and fixed $n \ne n'$ the terms run over the d distinct
roots of unity. (Note: b/2=b(d+1)/2 mod[d], as we consider modular numbers.) Thus we have a
unique solution, for some $m',$ given m, to
\begin{equation}\label{n}
\langle n|\hat{A}_{m,b}|n'\rangle=\langle n|\hat{A}_{m',b'}|n'\rangle,\;\;b\ne b',\;n\ne
n'.
\end{equation}
The  equality holds whenever, for fixed $n,n'\;n\ne n'$,
\begin{equation}\label{line1}
\frac{b}{2}\big(n+n'-1)-m=\frac{b'}{2}\big(n+n'-1)-m'.
\end{equation}
We now assert that that all the d projectors, one for each value of b, with fixed value of
$n+n'$ belong to a line. Adjuncted  by  the projector $|\ddot{m}\rangle\langle \ddot{m}|$
(that belong to the first column, $b=\ddot{0}$), with $2\ddot{m}=n+n'$, the set now forms
line. A convenient parametrization for the line obtains upon  rearranging Eq.(\ref{line1})
and taking $b'=0$ to get the equation for m as a function of b, i.e. the equation for the
line j=($\ddot{m},m(0))$, viz. Eq.(\ref{m(b)}).

 We now note that
projectors $\hat{A}_{m,b}$ that form the line share necessarily all the non diagonal matrix
elements $\langle s|\hat{A}_{m,b}|s'\rangle$ with $s+s'=\ddot{m}$ and all  the diagonal
elements (=$1/d$). This, while matrix elements not abiding with these requirements are
distinct. With this we may now evaluate the line operator for this underpinning scheme,
\cite{rev2,rev1}, noting that for this case, as noted above, the  balance formula
 Eq.(\ref{I}), is ${\cal{R}}=\Bbb{I}$. To illustrate these considerations
 we evaluate $\hat{P}_j,\;j=(\ddot{m}=1,m(0)=2)$. Via Eq.(\ref{m(b)}) and Eq.(\ref{line}) we
 have
\begin{equation}\label{p}
\hat{P}_j=\hat{A}_{(1,\ddot{0})}+\hat{A}_{(2,0)}+\hat{A}_{(1,1)}+\hat{A}_{(0,2)}-\Bbb{I}.
\end{equation}
Via Eq.(\ref{n}),
\begin{equation}
\hat{A}_{(1,\ddot{0})}=\begin{pmatrix}0&0&0\\0&1&0\\0&0&0\end{pmatrix},\;\hat{A}_{(2,0)}=\frac{1}{3}
\begin{pmatrix}1&\omega^2&\omega\\\omega&1&\omega^2\\\omega^2&\omega&1\end{pmatrix},\;\hat{A}_{(1,1)}=
\frac{1}{3}\begin{pmatrix}1&\omega&\omega\\\omega^2&1&1\\\omega^2&1&1\end{pmatrix},\;\hat{A}_{(0,2)}=
\frac{1}{3}\begin{pmatrix}1&1&\omega\\1&1&\omega\\\omega^2&\omega^2&1\end{pmatrix}.
\end{equation}
Eq.(\ref{p}) now gives
\begin{equation}
\hat{P}_j=\begin{pmatrix}0&0&\omega\\0&1&0\\\omega^2&0&0\end{pmatrix}.
\end{equation}
The general formula for the matrix elements of the line operator is
\begin{equation}\label{nxn}
\langle n|\hat{P}_{j=(\ddot{m},m(0))}|n'\rangle=\delta_{(n+n'),2\ddot{m}}\omega^{-(n-n')m(0)}.
\end{equation}
The proof of this is outlined in Appendix B \cite{rev1}.\\
This mapping of the Hilbet space projectors onto lines and points of the underpinning geometry was shown in \cite{rev2} to allow a convenient finite dimensional 
Radon trqansform.

\section{  Geometric Underpinning of Two-particles States}

We now consider DAPG underpinning for {\it states} of a $d^2$ dimensional {\it two}
particles, each of d-dimensional  Hilbert space. Our coordination scheme is as outlined
above $\alpha=(m,b);\;j=(\ddot{m},m(0)),\; m(b)=m(0)+b/2(2\ddot{m}-1)$. However, now each
point will refer to a two-particles {\it state} as is specified below. We have thus,

\begin{equation}
|A\rangle_{\alpha};\;\;\alpha= 1,2....d(d+1),\;\;|P_j\rangle;\;\; j=1,2,..d^2.
\end{equation}

$|A_{\alpha}\rangle$ are underpinned  with the d(d+1) points, $S_{\alpha}$ while the
$|P_j\rangle$ with the $d^2$ lines, $L_j$.\\
We define the  states , $|A_{\alpha}\rangle$,   underpinned by the geometrical points ,
 by

\begin{equation}\label{tilde}
|A_{\alpha}\rangle \equiv |m,b\rangle_1|\tilde{m},\tilde{b}\rangle_2.
\end{equation}
 $|\tilde{m},\tilde{b}\rangle$ is given by Eq.(\ref{cc}).

 With this we return to
states interrelation implied by the geometry: Eqs.(\ref{A}),(\ref{line}) now read
\begin{equation}\label{oprel3}
|A_{\alpha}\rangle=\frac{1}{d}\sum_{j\in
\alpha}^{d}|P_j\rangle\;\;\rightarrow\;\;|P_j\rangle=\sum_{\alpha \in
j}|A_{\alpha}\rangle\; -\;\sum_{\alpha'\in \alpha \cup M_{\alpha}}|A_{\alpha'}\rangle.
\end{equation}

( note that $|P_j\rangle$ is not normalized.) \\

We now show that with the choice , $\tilde{m}=d-m,\;\tilde{b}=d-b,$ the balance formula, viz
the base independence of the balancing term, Eq.(\ref{I}), holds: cf. \cite{fivel1,berge2},

\begin{eqnarray}
\sum_{\alpha{'}\in \alpha \cup M_{\alpha}}|A_{\alpha'}\rangle\equiv\sum_{m\in
b}^d|A_{m,b}\rangle= \sum_{m\in
b}^d|m,b\rangle_1|\tilde{m},\tilde{b}\rangle_2&=&\sum_{m,n,n'}^d|n\rangle_1|n'\rangle_2\langle n|m,b\rangle
\langle n'|\tilde{m},\tilde{b}\rangle \nonumber \\
=\sum_{n}^{d}|n\rangle_1|n\rangle_2&=&| {\cal{R}}\rangle\; \;\;independent\; of\;b\; \forall\;b.
\end{eqnarray}
This of course includes the first column, $b=\ddot{0},$ with the "point" in
the $n'$ row underpinning the state $|n'\rangle_1|n'\rangle_2$.\\

The relation among the matrix elements of projectors, $\hat{A}_{(m,b)}=|m,b\rangle\langle
b,m|,$ residing on the line given by  Eq.(\ref{m(b)}), \cite{rev1,rev2}, with the two particle states,
$|A_{(m,b)}=|m,b\rangle_1|\tilde{m},\tilde{b}\rangle_2,$ residing on the equivalent line,
Eq.(\ref{m(b)}), are now used to obtain an explicit formula for the line state,

\begin{eqnarray}\label{p1}
|P_{j=\ddot{m},m(0)}\rangle&=&\frac{1}{\sqrt d}\big(\sum_{m(b)\in j}|m,b\rangle_1|\tilde{m},\tilde{b}\rangle_2-|{\cal{R}}\rangle\big)= \nonumber \\
=\frac{1}{\sqrt d}\sum_{n,n'}|n \rangle_1|n'\rangle_2\big[\langle n|\sum_{m(b)\in j}\hat{A}_{m,b}\;-\;\Bbb{I}|n'\rangle\big]&=&\frac{1}{\sqrt d}\sum_{n,n'}|n \rangle_1|n'\rangle_2\delta_{n+n',2\ddot{m}}\omega^{-(n-n')m(0)}\;\;\forall\;b.
\end{eqnarray}
Where we used Eq(\ref{cc}) and  Eq.(\ref{nxn}).
 The expression for the line state will be put now in a more pliable form,\cite{fivel1},
\begin{eqnarray}\label{fivel}
|P_{j=\ddot{m},m(0)}\rangle&=&\frac{1}{\sqrt d}\sum_{n,n'}|n\rangle_1|n'\rangle_2\delta_{n+n',2\ddot{m}}\omega^{-(n-n')m(0)}= \nonumber \\
=\frac{\omega^{2\ddot{m}m(0)}}{\sqrt d}\sum_{n}|n\rangle_1|2\ddot{m}-n\rangle_2\omega^{-2nm(0)}&=&\frac{\omega^{2\ddot{m}m(0)}}{\sqrt d}\sum_{n}|n\rangle_1\hat{X}^{2\ddot{m}}_2 \hat{Z}^{2m(0)}_2{\cal{I}}_2|n\rangle_2 = \nonumber \\
&=&\frac{\omega^{2\ddot{m}m(0)}}{\sqrt d}\sum_{m}|m,b\rangle_1\hat{X}^{2\ddot{m}}_2 \hat{Z}^{2m(0)}_2{\cal{I}}_2
|\tilde{m},\tilde{b}\rangle_2.
\end{eqnarray}
The inversion operator ${\cal{I}}$ is defined via
${\cal{I}}|n\rangle=|-n\rangle=|d-n\rangle$. $\hat{X},\hat{Z}$ are defined in section II.
The orthonormality of $|P_j\rangle$ is proved in appendix C.

The central result of our geometrical underpinning is the following intuitively obvious overlap relation
\begin{eqnarray}\label{ovrlp}
\langle A_{\alpha=(m,b)}|P_{j=\ddot{m},m(0)}\rangle\equiv\langle
m,b|_1\langle\tilde{m},\tilde{b}|_2P_ {j=\ddot{m},m(0)}\rangle&=& \frac{1}{\sqrt
d}\delta_{m,\big(m(0)+b/2[2\ddot{m}-1]\big)},\;\;b\ne \ddot {0},\nonumber \\
\langle A_{\alpha=(n,\ddot{0})}|P_{j=\ddot{m},m(0)}\rangle\equiv\langle n|_1\langle n|_2 P_{j=\ddot{m},
m(0)}\rangle&=&\frac{1}{\sqrt d}\delta_{n,\ddot{m}},\;\;b=\ddot {0}, \;i.e.\;computational\;basis.
\end{eqnarray}\label{p2}
Thus the overlap of $|A_{\alpha=(m,b)}\rangle$ with $|P_j\rangle$ vanishes for $\alpha
\not\in j$ i.e. for $ m\ne m(0)+b/2(2\ddot{m}-1)$: Only if the point (m,b) is on the line j
the overlap is non zero. This is a remarkable attribute: Each and every one of the
observables $|m,b\rangle_1|\tilde{m},\tilde{b}\rangle_2\langle
\tilde{b},\tilde{m}|_2\langle b,m|_1$ has a {\it definite and known} value if measured in
the state $|P_{(\ddot{m},m(0))}\rangle$ yet its constituents single particle observables do
{\it not} commute. Indeed the single particle e.g. $|m,b\rangle_1\langle b,m|$  has a finite probability to be found anywhere  (on every line). The probability of finding our system in the state $|A_{\alpha}\rangle$
given that the system is in the state $|P_j\rangle,\;\alpha\in j$ is  $\frac{1}{d}$.
We note, however, that there are d+1 points $\alpha$, exposing   that these probabilities
are not mutually exclusive. This can be directly checked  by noting the non vanishing of
the overlap, $|\langle A_{\alpha}|A_{{\alpha}'}\rangle|=1/d,\;\alpha\ne{\alpha}',\;\alpha,
{\alpha}'\in j.$ The probability when $\alpha \not\in j$ is nil. This  allows  a new
approach to the Mean King Problem to which we shall turn after the collective coordinate
formulation.

\section {Geometric View of collective Coordinates formulation}

The simplification offered by the collective formulation is illustrated by considering the balance term, cf. Eq(\ref{R}),  including normalization, Eq.(\ref{p1}),
\begin{equation}
\frac{1}{\sqrt d}|{\cal{R}}\rangle=\frac{1}{\sqrt d}\sum_n^d|n\rangle_1|n\rangle_2=\frac{1}{\sqrt d}\sum_n
\frac{1}{\sqrt d}\sum_{{n}',n"}|{n}'\rangle_r|n"\rangle_c\langle{n}'_r,n"_c|n\rangle_1|n\rangle_2
=\sum_n|0\rangle_r|n\rangle_c=|0;\ddot{0}\rangle_r|0;0\rangle_c.
\end{equation}
We used Eq.(\ref{relcom})to get $n'_r=0,\;n"_c=0$. The RHS reads that within the collective
coordinates the state $|{\cal{R}}\rangle$ is a product state: In the r (relative)
coordinates it is in computational basis ($b_r=\ddot{0}$) with eigenvalue (of $\bar{Z}_r$) 1
(i.e. m=0). In the c (center of mass)
coordinate space it is in $b_c=0$  with eigenvalue (of $\bar{X}_c$) 1 too.\\
We now turn to the expression for $|P_j\rangle$ within the collective coordinates system, using Eq.(\ref{p1}),
\begin{eqnarray}\label{pj}
|P_{j=\ddot{m},m(0)}\rangle&=&\frac{1}{\sqrt d}\sum_{n,n'}|n \rangle_1|n'\rangle_2\delta_{n+n',2\ddot{m}}
\omega^{-(n-n')m(0)}\nonumber \\
&=&\frac{1}{\sqrt d}\sum_{n,n'}\sum_{{n'}_r,{n"}_c}|n'\rangle_r|n"\rangle_c\langle{n'}_r,{n"}_c|n_1,n'_2
\rangle
\delta_{n+n',2\ddot{m}}\omega^{-(n-n')m(0)}\nonumber \\
&=&\frac{1}{\sqrt d}\sum_{n_r,n_c}|n\rangle_c|n'\rangle_r\delta_{n_c,\ddot{m}}\omega^{-2n_rm(0)}=
|m;\ddot{0}\rangle_c|2m(0);0\rangle_r.
\end{eqnarray}
Identifying the state as a product state in the collective coordinates. (The product state above is notationally  simplifid by $|\ddot{m}\rangle_c|2m_0\rangle_r.$)\\
The collective coordinate expression for the particles product  state $|A_{\alpha=(m,b)}\rangle$ is,
\begin{eqnarray}
|A_{\alpha=(m,b)}\rangle&=&|m,b\rangle_1|\tilde{m};\tilde{b}\rangle_2=\frac{1}{d}\sum_{n,n'}|n\rangle_1
|n'\rangle_2\omega^{(n-n')[\frac{b}{2}(n+n'-1)-m]},\nonumber \\
&=&\sum_{k,k'}|k\rangle_r|k'\rangle_c \omega^{2k[\frac{b}{2}(2k'-1)-m]}.
\end{eqnarray}
The probability amplitude of finding the particles in the state  $|A_{\alpha=(m,b)}\rangle$ given a system in the state  $|\ddot{m};\ddot{0}\rangle_c|2m(0);0\rangle_r\equiv |\ddot{m}\rangle_c|m_0\rangle_r$ is
\begin{eqnarray}
\frac{1}{d}\sum_{k,k'}\langle k|2m(0);0\rangle \langle k'|\ddot{m}\rangle
\omega^{2k[\frac{b}{2}(2k'-1)-m]} &=&\frac{1}{d^{\frac{3}{2}}}\sum_{k}\sum_n \langle
k|n\rangle \omega^{-2nm(0)}\omega^{2k[\frac{b}{2}(2\ddot{m}-1)-m]}=
\frac{\delta_{m,(m(0)+\frac{b}{2}(2\ddot{m}-1))}}{\sqrt d};\;b\ne\ddot{0}\nonumber \\.
\langle n|_1\langle n|_{2} 2m(0);0\rangle_r|\ddot{m}\rangle_c&=&\frac{\delta_{n,\ddot{m}}}{\sqrt
d},\;b=\ddot{0}.
\end{eqnarray}
Thus the {\it probability} is $\frac{1}{d}$ if the state is on the line (nil if it is not),
confirming
 Eq.(\ref{ovrlp}) and the efficiency of the collective coordinate formulation.\\

\section{Leaky  Particles}

The maximally entangled state, Eq.(\ref{pj}),  $$|P_{j=(\ddot{m},m_0)}\rangle\equiv |\ddot{m}\rangle_c|2m_0\rangle_r,$$ was viewed as a "line" state.  I.e.  the product states  underpinned by the geometrical point, $\alpha=(m,b),$
$$|A_{\alpha}\rangle=|m,b\rangle_1|\tilde{m},\tilde{b}\rangle_2,$$
whose coordinates, (m,b),  abides by the line equation, Eq. (\ref{m(b)}),
$$m(b)=m_0+\frac{b}{2}(2\ddot{m}-1),\;\;b\ne \ddot{0},\;\;m(\ddot{0})=\ddot{m}$$
form form line in the sense that ( cf.  Eq.(\ref{p2})),
\begin{eqnarray}
\langle A_{\alpha}|P_j\rangle \equiv \langle\tilde{b},\tilde{m}|_2\langle b,m|_1\ddot{m}\rangle_c|2m_0\rangle_r&=&\frac{1}{d},\;\;\alpha \in j,\ nonumber \\
                                                                                                                                                                                   &=& 0 \;\;\;\alpha \not\in j.
\end{eqnarray}

Thus a pair of particles (the particle and its mate, the tilde particle) whose coordinates are $\alpha=m,b$ do wholy belong to the d  lines that share the coordinated point.
However each of the constituent particles (either 1 or 2) is {\it equaly} likely to be in anyone the $d^2$  of the lines,

\begin{equation} \label{mbj}
\langle b,m|_1 \ddot{m}\rangle_c|2m_0\rangle =\frac{1}{\sqrt d}|(\tilde{m}-2\tilde{\Delta}),\tilde{b}\rangle_2\omega^{2\ddot{m}\Delta},\;\;\;\Delta=m-m_0+\frac{b}{2}(2\ddot{m}-1),
\end{equation}
($\tilde{\Delta}=d-\Delta.$)   Thence,

$$|\langle b,m|_1\ddot{m}\rangle_c|2m_0\rangle|^2=\frac{1}{d},\;\;\forall (\ddot{m},m_0).$$ 
 It is this attribute that allows the tracking of the King measurement alignment.

\section{Tracking the  Mean King }

The Mean King Problem (MKP), initiated by \cite{lev1},  was analyzed in several
publications - see the comprehensive list in \cite{berge2}. Briefly summarized it runs as
follows. Alice may prepare a state to her liking. The King measures it in an MUB basis (i.e. for some value of b: a particular alignment of his apparatus). He
does not inform Alice of his observational result nor the basis he used. 
 Alice performs a control measurement of her choice. {\it After} her control measurement the King informs her the basis, b, he used for his measurement. Thence she must {\it deduce} the actual state (m,b) that he observed.  In our case of {\it tracking} the King - He does not inform Alice of the basis he used - her control measurment  is designed to track the basis used. (Note that in all the analyses time evolution is ignored - presumed to be independently accountable.)\\

  The state that Alice prepars  is one of the line vectors,
$$|P_{j=(\ddot{m},m(0))}\rangle=|\ddot{m}\rangle_c|2m_0\rangle_r$$.
 Thus she knows both $\ddot{m}$ and m(0). The King's
measurement is along a line of some fixed b i.e. he measures
$$\sum_m|m,b\rangle \omega^m \langle b,m|$$
and observed, say, $\omega^m$. The Kings measurement projects the state $|P_j\rangle$ to (neglecting normalization):

$$|m,b\rangle_1 \langle b,m|_1|\ddot{m}\rangle_c|2m_0\rangle_r.$$
Now Alice measures the {\it non degenerate} operator,
$$\sum_{m',m"}|\ddot{m}{'}\rangle_c|2m_0^"\rangle_r \gamma_{m',m"}\langle 2m_0^"|\langle \ddot{m}{'}|,$$
and obtains, say, $\gamma_{m',m"}$. Thence the quantity,
\begin{equation}\label{b}
\langle 2m_0^"|_r\langle \ddot{m}{'}|_c m,b\rangle_1 \langle
b,m|_1|\ddot{m}\rangle_c|2m_0\rangle_r\ne 0.
\end{equation}
The LHS of this equation is obtained by using Eq.(\ref{mbj}) and noting that 
$$\langle 2m_0^"|_r\langle \ddot{m}{'}|_c m,b\rangle_1$$
may be obtained from it by replacing $\ddot{m},m_0$ by $\ddot{m}{'},m_0^{'}$
and taking the complex conjugate expression. The result is,
\begin{equation}
\langle 2m_0^"|_r\langle \ddot{m}{'}|_c m,b\rangle_1 \langle
 b,m|_1|\ddot{m}\rangle_c|2m_0\rangle_r=\frac{1}{d}\delta{[(m_0^{'}-m_0),b(\ddot{m}-\ddot{m}{'})]}.
\end{equation}
i.e.
\begin{equation}
b=\frac{(m_0^{'}-m_0)}{(\ddot{m}-\ddot{m}{'})};\;\;\;\ddot{m}=\ddot{m}{'}\rightarrow b=\ddot{0}.
\end{equation}

Knowing the initial state, i.e. $\ddot{m}$ and $m_0$ and measuring the final state, viz. $\ddot{m}{'}$
 and $m_0{'}$ Alice tracks b - the apparatus alignment used by the King. (The alignment $b=\ddot{0}$ - the King's CB - gives $\delta_{\ddot{m},\ddot{m}{'}}.$)

\section {   Summary and Concluding Remarks}
We outlined finite plane geometrical underpinning to {\it states} of two d-dimensional particles Hilbert spaces. d=prime ($\ne 2$).
 The geometrical
points $\alpha$,  were coordinated with mutual unbiased bases (MUB) states labels: $\alpha=(m,b)$.
m denotes the vector in the base b, it  locates the vertical coordinate within the column
labelled by b. b gives the position of the column. Points of the geometry $\alpha=(m,b)$
underpin product states labelled with (m,b):
$|A_{\alpha}\rangle=|m,b\rangle_1|\tilde{m},\tilde{b}\rangle_2,$ 1 and 2 refers to the
particles with $\tilde{m}=d-m,\;\tilde{b}=d-b.$  A geometrical line, j,  runs through d+1 points (one point at each MUB basis) - it underpins the state
 vector $|P_{j=(\ddot{m},m_0)}\rangle$ . This vector is completely parametrized by two points $(m(\ddot{0}),m(0))\equiv(\ddot{m},m_0)$ - the point on the computational basis column
$(b=\ddot{0})$  and the point on the b=0 column i.e.  the eigenfunctions of the displacement oparator.   Viewing a
line as a point's evolution with increasing b, l \cite{svet},  the line  label specified with $\ddot{m}$
and m(0) may be viewed as specification in terms of initial position and momentum \cite{torre}.\\
 The states $|P_j\rangle,\;\;j=1,2,...d^2$  are  maximally entangled  and form an orthnormal basis that spans the space. 
These states  have a remarkable attribute: the probability of finding it in the state $|A_{\alpha}\rangle$,
$\alpha \in j$ is 1/d, it vanishes otherwise. The number of points ,$\alpha$, on a line is d+1 -  reflecting the {\it non}-exclusiveness of these probabilities
which , in turn, allows the tracking of the King measurement  as is accounted above.\\
We gave an alternative , perhaps more economic, parametrization of the $d^2$ {\it  maximally entangled}  states that span the space - parametrization based on a collective, 
viz center of mass and realtive, coordinates. Here the state vector underpined with a geometrical line is given by a {\it product} state of the collective coordinates,
$|P_{j=\ddot{m},m_0}\rangle=|\ddot{m}\rangle_c|2m_0\rangle_r$  (c and r are center of mass and realtive coordinattes respectively). These states were shown to  
provide simplified noation for the calculation as well as a novel view of maximally entangled states.\\

It was shown that adding up product states in geometrically reasoned manner yields maximally entangled states. These states are shown to be product states of two particles collective (center of mass and realtive) coordinates. The states are such as to allow unambiguous tracking  of alignment of  measurement  of their constituent single particle.\\

\section*[Appendix A] {Appendix A: Geometrically based Hilbert space operators' interrelation}

Our task is to define consistently addition (and subtractions)  of  " line" and "point"  Hilbert space  operators  (or states)  which are   underpinned by geometrical points and lines   assuring that they abide by their geometrical underpinning interrelation. The logical  interrelation symbols (S and L represents the geometrical point and line respectively),
$$\bigcup_{\alpha \in j}^{d}S_{\alpha}=L_j;\;\;\;\bigcap_{j \in \alpha}^d L_j=S_{\alpha}$$
are to be realized by addition ( and subtraction)  of Hilbert space entities, operators or states,  supplamented with numerical values.  Our starting point is :

\begin{equation} \label{a}
S_{\alpha}=\bigcap_{j \in \alpha}L_j \;\;\Rightarrow \;\;A_{\alpha}=\frac{1}{d}\sum_{j \in \alpha} P_j,
\end{equation}

where we underpinned  the Hilbert space operator (or state) $A_{\alpha}$ with the point $S_{\alpha}$. We now consider a particular realization of the geometry, i.e.  a set up where the points and lines abide by the geometry are realized by  marking  the points on each line subject to DAPG requirements such as, e.g., two distinct lines have a single point in common. The geometry is then realized  via Eq.(\ref{a}), coordinated as specified by MUB labelings. It, then, follows via DAPG(a,c,d,e) - cf. Eq.(\ref{R}), that

\begin{equation}
\sum_{\alpha{'} \in \alpha \cup M_{\alpha}}  A_{\alpha{'}}=\frac{1}{d}\sum_{\alpha{'} \in \alpha \cup M_{\alpha}}\big( \sum_{j \in \alpha{'}} P_j\big)=
\frac{1}{d}\sum_{j=1}^{d^2}P_j.
\end{equation}

The RHS is clearly a universal quantity (i.e. independent of $\alpha$ and j) which implies that the LHS,

$$\sum_{\alpha \in \alpha' \cup M_{\alpha'}}A_{\alpha}= {\cal{R}},\;\;independent\; of \;\;\alpha',$$
i.e. universal too.\\

 Since a line is made of points we consider (try)
\begin{equation}\label{b}
P_j=\sum_{\alpha \in j}A_{\alpha} -{ \cal{R}},
\end{equation}
where ${\cal{R}}$ is a universal quantity  that may be required to balance the equation. Returning to Eq.(\ref{a}) with Eq. (\ref{b}), {\it the geometry} 
implies via DAPG(c,d),

\begin{equation}
A_{\alpha}=A_{\alpha}+d\sum_{\alpha \in \alpha' \cup M_{\alpha'}} A_{\alpha}-  d {\cal{R}};\;\;\Rightarrow\;\;\sum_{\alpha \in \alpha' \cup M_{\alpha'}}A_{\alpha}=  {\cal{R}}.
\end{equation}

We illustrate the consistency of this by showing the validity of the geometrically derived realtion, Eq.(\ref{I}):

$$\frac{1}{d}\sum_j^{d^2} P_j=\frac{1}{d}\sum_j^{d^2}\big(\sum_{\alpha \in j}^{d+1} A_{\alpha} -{\cal{R}}\big)=-d{\cal{R}}+\sum_{\alpha}^{d(d+1)} A_{\alpha}=$$
$$=-d{\cal{R}}+(d+1){\sum_{\alpha{'} \in \alpha \cup M_{\alpha}}^{d}  A_{\alpha{'}}}={\cal{R}}. \;\;QED.$$
Where we used the universality of ${\cal{R}}$ and DAPG(c,d).
Thus 
$$P_j=\sum_{\alpha \in j}A_{\alpha}-{\cal{R}}.$$
With, 
\begin{eqnarray}
A_{\alpha}&=&|m,b\rangle \langle b,m|\;\;\Rightarrow \;\;{\hat{\cal{R}}}={\hat{\Bbb{I}}}; \nonumber \\
A_{\alpha}&=&|m,b\rangle_1|\tilde{m},\tilde{b}\rangle_2\;\;\Rightarrow \;\;|{\cal{R}}\rangle= \sum_{m}|m,b\rangle_1|\tilde{m},\tilde{b}\rangle_2.
\end{eqnarray}

Distintict consistent constructions are possible,\cite{tomer},  for example with a starting point,
\begin{equation}
\hat{P}_j=\frac{1}{d+1}\sum_{\alpha \in j}^{d+1}\hat{A}_{\alpha},
\end{equation}
one is lead to
$$\hat{A}_{\alpha}=\frac{d+1}{d}\sum_{j \in \alpha}^d \hat{P}_j-{\hat{\cal{R}}}.$$

\section*[Appendix B] {Appendix B: The Line operator $\hat{P}$}

With the geometrical point, $\alpha=(m,b)$, underpinning the MUB {\it projector} $\hat{A}_{\alpha=m,b}=|m,b\rangle\langle b,m|$ the geometrical line 
$j=(\ddot{m},m_0)$ underpins the operator $\hat{P}_j$, Eq. (\ref{nn'},\ref{line}, \ref{p}),
$$\hat{P}_j=\sum_{\alpha\in j} \hat{A}_{\alpha} -\Bbb{I}.\;\;\;\;\;\langle n|\hat{A}_{\alpha=(m,b)}|n'\rangle=\frac{\omega^{(n-n')[\frac{b}{2}(n+n'-1)-m]}}{d}.$$

A line ,$\hat{P}_j$ is given by the contribution of point projectors $\hat{A}_{\alpha}$ with common matrix elements, i.e. $\hat{A}_{\alpha}$ and $\hat{A}_{\alpha{'}}$  belong to the same line  whenever $\langle n|\hat{A}_{\alpha}|n'\rangle=\langle n|\hat{A}_{\alpha{'}}|n'\rangle$  ( for b$\ne \ddot{0}$) . This reads for n+n'=common constant which we chose to be
$2\ddot{m}$ - which a point on the line on the $b=\ddot{0}$ column. The next point on the line is the value of m on the b=0 column, $m(0)=m_0$ . The line is now defined by the two points
$j=(\ddot{m},m_0)$. All other points are now given by 
$$ -m(0)=\frac{b}{2}(2\ddot{m}-1)-m(b)\;\rightarrow  m(b)=\frac{b}{2}(2\ddot{m}-1)+m_0.$$
Since alll theother matrix elements of the projectors forming the line are distinct and are, each, a d-root of unity their sum add up to zero. Thus the final formula for $\hat{P}_j$ is
$$\langle n|\hat{P}_j|n' \rangle=\delta_{(n+n'),2\ddot{m}} \omega^{-(n-n')m_0}.\;\;\;QED$$

\section*[Appendix C] {Appendix C: Orthogonality of $|P_j\rangle$}

Noting that $\langle{\cal{R}}|{\cal{R}}\rangle=d$ and $\langle P_j|{\cal{R}}\rangle=d+1$,
We get, for j=j':

\begin{equation}
\langle P_j|P_j\rangle= \frac{1}{d}\{\sum_{\alpha \in j}\langle
A_{\alpha}|-\langle{\cal{R}|\}}\{\sum_{\alpha' \in
j}|A_{\alpha'}\rangle-|{\cal{R}}\rangle\}=\frac{1}{d}\{1+d +\sum_{\alpha\ne
\alpha'}^{d+1}[\langle A_{\alpha}|A_{\alpha'}\rangle]-2(d+1)+d\}=1
\end{equation}
Where we used that $\sum_{\alpha\ne\alpha'}^{d+1}[\langle
A_{\alpha}|A_{\alpha'}\rangle=(d+1)\frac{d}{d}=d+1$.

For $j\ne j'$ the geometry dictates, DAPG(b), that distinct lines share {\it one} point.
Thus the first term above is 1 rather than 1+d hence
$$\langle P_j|P_{j'} \rangle=0, \;\;j \ne j'$$
i.e.
\begin{equation}
\langle P_j|P_{j'}\rangle=\delta_{j,j'}.\;\;QED \\
\end{equation}

Note: Using the collective coordinates , $|P_{j=(\ddot{m},m(0))}\rangle=|\ddot{m}\rangle_c|m_0\rangle_r$,  the proof
is immediate.\\

Acknowledgments: The hospitality of the Perimeter Institute where this work was completed and informative discussions with Prof. L. Hardy, A. Mann, P.A. Mello and M. Mueller are gratefully acknowledged.

\end{document}